\documentclass[11pt]{article}

\usepackage{amsmath}
\usepackage{amssymb}

\newcommand{\beq}{\begin{equation}}
\newcommand{\eeq}{\end{equation}}
\newcommand{\bit}{\begin{itemize}}
\newcommand{\eit}{\end{itemize}}
\newcommand{\ben}{\begin{enumerate}}
\newcommand{\een}{\end{enumerate}}

\newcommand{\ch}{{\cal{H}}}

\newcommand{\bc}{{\boldsymbol{C}}}
\newcommand{\cu}{{\cal{U}}}
\newcommand{\fh}{{\mathfrak{H}}}
\newcommand{\cm}{{\cal{M}}}
\newtheorem{Theorem}{Theorem}

\newcommand{\bs}{\boldsymbol}

\begin{document}

\begin{titlepage}

\begin{flushright}
\today
\end{flushright}

\vspace{1in}

\begin{center}

{\bf D-particle Field Category}

\vspace{1in}

\normalsize

{Eiji Konishi\footnote{E-mail address: konishi.eiji@s04.mbox.media.kyoto-u.ac.jp}}

\normalsize
\vspace{.5in}

 {\em Faculty of Science, Kyoto University, Kyoto 606-8502, Japan}

\end{center}

\vspace{1in}

\baselineskip=24pt
\begin{abstract}
We formulate the homotopy associative ($A_\infty$) category of D-particle field states via gauged S-duality.
By invoking the minimal model theorem of this D-particle field category, we investigate the equivalence principle and the $A_\infty$ covariance principle in the theory of gauged S-duality considered as D-particle field theory.
\end{abstract}

\vspace{.7in}
 
\end{titlepage}
\section{Introduction}

This paper is based on the author's previous paper\cite{Konishi} in which we have given a basic formulation of a theory of gauged and affinized S-duality\cite{Konishi,Sen,Schwarz} that describes type IIB string theory vacua.
Here, we investigate the equivalence and homotopy associative ($A_\infty$) covariance principles in the theory of gauged S-duality by considering the geometry of the D-particle (D-0 brane) field theory corresponding to type~IIA/M-theory.\cite{Witten,Polchinski,Y1,Y2,Y3,K}
There are several approaches to the constructive definition of superstrings and M-theory.\cite{BFSS,IKKT} In particular, it has been recognized that they can be considered as second quantized theories of D-branes. Therefore, our present investigation aims to provide the basis for the geometry of a constructive definition of string theory.
Different constructive definitions of D-particle fields have been proposed by Yoneya and the author.\cite{Y1,Y2,Y3,K} In particular, the author has defined D-brane field theory by using a single ${\widehat{sl(2,{\bs{R}})}}$ Becchi-Rouet-Stora-Tyutin (BRST) charge $Q$ and the kernel of its wave functions \begin{equation}Q\Psi^Q[{g},\vec{s}]=0\;,\label{eq:KO}\end{equation}
where ${g}$ is the modulus parameter of fundamental string world sheets and $\vec{s}=(s_n)_n$ are the coordinates on the infinite dimensional base space of the ${\widehat{sl(2,{\bs{R}})}}$ fiber bundle.
This wave function $\Psi^Q$ is {{not that of matter but that of the Universe}}. As a result, there is no fixed time parametrization of $\Psi^Q$ itself.
The theory of gauged and affinized S-duality can then be formulated as an ${\widehat{sl(2,{\bs{R}})}}$ Yang-Mills theory with an infinite number of time variables.\cite{Konishi}

The condition in Eq.(\ref{eq:KO}) is generalized by introducing a nonlinear (second gauge) potential and defining a covariant derivative $\nabla$ (when we regard the BRST charge as a differential) to produce a condition on the sections in the vector bundle parallel to $\nabla$:\cite{Konishi}
\begin{equation}
\nabla \Psi^\nabla[{g},\vec{s}]=0\;.
\end{equation}
The equation of a single vanishing curvature which defines the covariant derivatives vanishes under the action of the covariant derivative due to the generalized gauge invariance.\cite{Konishi}

In this paper, we describe the geometry of the motion of D-particles and open strings in a Universe with wave function $\Psi^Q$ in the language of the second quantized generalization of the quantum cohomology of the field theory of fundamental strings in the sense of the second quantization of D-particle fields. The wave function $\Psi^Q$ is recognized as the local description of the wave function $\Psi^\nabla$, promoted by including an infinite number of time coordinates; although, as already mentioned, for $\Psi^Q$ there is no fixed time parametrization.
 In the following geometrical formulation, we identify the D-particles with geodesics on the upper half plane. (The basis for this identification will be explained in the next section.)
 We study the homotopy structure of the second quantized D-particle field theory that includes the degrees of freedom of multiple D-particles with Chan-Paton factors on them, and we construct the $A_\infty$ category $C^Q({\mathfrak{H}})$ of background independent D-particle field states by considering the dynamics of multiple D-particles on the upper half plane ${\mathfrak{H}}$ of the string coupling constant $g_s$. {{The $SL(2,{\bs{R}})$ BRST transformations promote the geodesics to other geodesics on the upper half plane ${\mathfrak{H}}$ given by the coordinate $s_n$, that is, the hidden time variables}}, since {{the Poincar\'e metric on the upper half plane is $SL(2,{\bs{R}})$ invariant}}.
    Then, our new viewpoint is to consider the dynamics of multiple D-particles, which are mapped onto a set of geodesics on the upper half plane of ${g_s}$ deformed by an infinite number of coordinates. Here, as detailed in next section, we rule that D-particles are superposed only if their corresponding geodesics intersect on $\fh$.
The Feynman diagram of multiple (anti-)D-particles with bounded open string fields on ${\mathfrak{H}}$ contains the intersection of each set of geodesics on the upper half plane, for example, $n$-multiplicity D-particles and $m$-multiplicity D-particles fuse into $m+n$-multiplicity D-particles on ${\mathfrak{H}}$. (Here, we count the multiplicities of anti-D-particles by using negative integers.)
The proper language of the homotopy product structure for bounded open string fields is an $A_\infty$ structure. 
We invoke the minimal model theorem which ensures the existence of the quasi-isomorphism $A_\infty$ functor ${\cal{U}}$ between two $A_\infty$ categories, $C^Q({\mathfrak{H}})$ and its minimal model $H(C^Q(\fh))$, which secondarily results in the connection between the D-particle field state spaces of $\Psi^\nabla$ for systems $S_1$ and $S_2$ defined in terms of local coordinates:
\begin{equation}
{\cal{U}}:V^\nabla|_{S_1}\to V^\nabla|_{S_2}\;.\label{eq:trf}
\end{equation} 
On the basis of this $A_\infty$ functor, the equivalence principle for all the quantized Chan-Paton gauge interactions from the point of view of gauged S-duality is formulated by invoking the minimal model theorem,\cite{Kad,Kajiura} and we suggest that $A_\infty$ covariance is a principle of the theory of gauged S-duality.

\section{D-particle Field Category: A Toy Model}
An $A_\infty$ category $C$ is the triple of a set of objects $C=\{\bc_i\}_i$, the morphism spaces between two arbitrary objects ${\cal{H}}(\bc_i,\bc_j)$, which are ${\bs{Z}}$-graded vector spaces, and the product structures $m_k$ between $k$ morphism spaces among $k+1$ arbitrary objects $\bc_i$ for $1\le i\le k+1$. The product structures are degree $(2-k)$ multi-linear maps
\begin{equation}
m_k:\bigotimes_{i=1}^k{\cal{H}}(\bc_i,\bc_{i+1})\to{\cal{H}}(\bc_1,\bc_{k+1})\;,\end{equation}
for $k=1,2,\ldots$, which satisfy the $A_\infty$ conditions (see Eq.(\ref{eq:l})).\cite{Fukaya}

We define an $A_\infty$ category such that its object set is the set of the sets of geodesics $C_1,C_2,\cdots$ promoted to other geodesics by an infinite number of coordinates $\vec{s}$ on the Poincar\'e upper half plane ${\mathfrak{H}}$:
\begin{equation}{C^Q({\mathfrak{H}})}=\bigl\{ \bc_i\bigr\}_i\;,\ \ \bc_i=\{C_{i_j}\}_j\;.\end{equation}

These sets of geodesics $\bc$ represent the background independent (multiple) D-particle field states. (When S-duality is gauged, the distinction between D-particles and fundamental strings is removed. On this point, as stated in the previous paper,\cite{Konishi} we assign Ramond-Ramond (R-R) and Neveu-Schwarz-Neveu-Schwarz (NS-NS) parts of states on the basis of $\widehat{sl(2,{\bs{R}})}$ generators.) We introduce the morphism structure between the objects by invoking that of the Fukaya category.\cite{Fukaya} We note that geodesics in the upper half plane represent parts of the moduli spaces of S-duality doublets obtained by fixing the R-R sectors, which run along the real axis of the upper half plane, and by keeping the degrees of freedom of NS-NS sectors. Thus, the intersections between geodesics represent the situation in which fundamental strings (an NS-NS sector) connect D-particles (an R-R sector). We define D-particle multiplicities based on these arguments using geodesics. We recognize two geodesics that do not intersect in $\fh$ but {{only at $i\infty$ or $\partial\fh$}} as non-intersecting D-branes; in the D-particle case these are two single (not multiple) D-particles. We note that for any set of D-branes, the open string field degrees of freedom connecting them exist. The fact that geodesics in $\fh$ always intersect at least in the boundary of $\fh$ (including infinity) reflects this fact. Therefore, we introduce the class of geodesics by the condition that when $C_1$ and $C_2$ intersect on $\fh$ we consider them to be equivalent, $C_1\sim C_2$. This equivalence relationship gives a class by the fact that if $C_1\sim C_2$ and $C_2\sim C_3$ then $C_1\sim C_3$ (i.e., transitivity). We consider the number of geodesics $n$ that belong to the same class as multiple (anti-)D-particles with multiplicity $n$. Hence, the open string field state space is the BRST cohomology of $U(|n|)$ Chan-Paton gauge theory $Q_n$.
If we denote the intersection with $n$-multiplicity, which is defined by the number of elements in the equivalence class to which $\alpha$ belongs, by $\alpha^n$ (e.g., if $C_1\cap C_2=\{i\infty\}$ then $n=1$), the morphism space is defined by the ${\boldsymbol{Z}}$-graded vector space
 \begin{eqnarray}
\ch(\bc_i,\bc_j)&=&\bigoplus_k\left\{\begin{array}{c|c}&i\hbar\frac{\delta \alpha_k(\vec{s})}{\delta \tau(\vec{s})}=Q(\vec{s})|_{V_{n_k}}\alpha_k(\vec{s})\\ \alpha_k(\vec{s})&\\ &\ \ \alpha_k(\vec{s}_{{in}})=\bc_i\;,\ \ \alpha_k(\vec{s}_{{f}})=\bc_j\end{array}\right\}\;,\nonumber\\
 \bc_i\cap \bc_j&=&\bigoplus_k\alpha_k^{n_k}\;,\ \ n=\sum_kn_k\;,
 \label{eq:mor}
 \end{eqnarray}
 for an integer $n$ and the initial and final coordinates $\vec{s}_{in}$ and $\vec{s}_f$ of each flow $\alpha$, where $Q(\vec{s})$ is the infinitesimal $SL(2,{\bs{R}})$ BRST transformation (the parameter ${g}$ is fixed), $\delta\tau(\vec{s})$ is the increment of the cosmic time\cite{Konishi} and $V_{n}$ is the BRST cohomology vector space of BRST charge $Q_{n}$. Each morphism has the variables of the representation space\cite{Konishi}, and these variables parametrize the geodesics. The ${\boldsymbol{Z}}$ degree of each morphism is given by the (anti-)D-particle multiplicity $n$. We simplify the BRST cohomology of $Q_n$ to be the cohomology numbers of the space of the $n$-forms on moduli spaces that will be introduced later. The derivative operator of these differential forms is the BRST charge $Q$. We note that the process of intersection between geodesics is described by the degeneration process in the infinities of the moduli space of the geodesics, which correspond to the definition of the objects.

 We make two remarks about Eq.(\ref{eq:mor}). First, D-particles do not intersect with each other. Only the interactions between D-particles promote the open string fields. So $\alpha(\vec{s})$ includes the degrees of freedom of D-particles as well as those of the open string fields. Second, in the previous paper, we defined the objects of the category $C^Q$ by the complete set of the subspaces in the full Hilbert space.\cite{Konishi} These subspaces correspond to the static geodesics in our definition of the objects of $C^Q(\fh)$.

We introduce the glue of two flows $\alpha_1$ and $\alpha_2$ by\cite{planar}
\begin{equation}
\alpha_{ij}\star\alpha_{jk}=\sum_{v} {\mbox{sgn}}(\alpha_{ik})S_2({\vec{\alpha}}(v))\alpha_{ik}\;,
\end{equation}
where the definition of ${\mbox{sgn}}$ is same as that in Ref.15. The production amplitude $S_2$ will be defined soon (see Eq.(\ref{eq:l1})) and each vertex $v=v_{ijk}$ between two flows is defined by $v_{ijk}=\alpha_{ij}(\vec{s}_m)=\alpha_{jk}(\vec{s}_m)$ for its certain medium coordinates $\vec{s}_m$.

It is natural from a physical point of view to assume that the higher product structures of $\alpha_{i,j}$ in ${\cal{H}}({\bc_i},{\bc_j})$ between $k$ ($k=2,3,\ldots$) sets of geodesics are defined as the scattering amplitudes of open string fields. In our toy model, the product structures on {$C^Q(\mathfrak{H})$} are defined by
\begin{eqnarray}{m}^Q_1(\alpha)(x)&=&Q\alpha(x)\;,\label{eq:low}\\
(m^Q_k(\alpha_1,\cdots,\alpha_k))(x)&=&\sum_{\alpha_{k+1}=\alpha_1\star \alpha_2\star\cdots \star\alpha_k}\sum_v{\mbox{sgn}}(\alpha_{k+1})S_k({\vec{\alpha}}(v))\alpha_{k+1}(x)\;,
\end{eqnarray}where we introduce the production amplitudes\cite{PA}
\begin{eqnarray}
S_k(\alpha_1,\cdots,\alpha_k)=\int_{({\boldsymbol{x}}^k)\in\cm_{\alpha_1,\alpha_2,\ldots,\alpha_k}}\alpha_1(x_1)\wedge\alpha_2(x_2)\wedge\cdots\alpha_k(x_k)\;,\label{eq:l1}
\end{eqnarray}
and we put $\alpha_i=\alpha_{i,i+1}$ for $1\le i\le k$ and $\alpha_{k+1}=\alpha_{1,k+1}$.
The integral domain of Eq.(\ref{eq:l1}), that is, the moduli space $\cm_{\alpha_1,\alpha_2,\ldots,\alpha_k}$ satisfies the cyclic symmetry
\begin{equation}
{{\cm}}_{\alpha_1,\alpha_2,\ldots,\alpha_k}=(-)^{k+1} {{\cm}}_{\alpha_{2},\alpha_{3},\ldots ,\alpha_{1}}\;,\label{eq:cyc1}
\end{equation}
where the sign represents the orientation, and we define its coboundary operation by
\begin{eqnarray}
\partial\cm_{\alpha_1,\alpha_2,\ldots,\alpha_n}=&&\bigcup_{k=1}^n\bigcup_{l=1}^{n-3}\Bigl((-)^{(n+1)(k+l+1)}\times\nonumber\\ &&\times(\cm_{\alpha_{k},\ldots,\alpha_{k+l},\alpha_a}\times\cm_{\alpha_{a^\prime},\alpha_{k+l+1},\ldots,\alpha_{k+n-1}})\Bigr)\;,\label{eq:cyc2}\end{eqnarray}
which satisfies $\partial^2=0$\cite{Nakatsu} in the dual concept, that is, the integral domain, of the forms. 

Nilpotency for $C^Q(\mathfrak{H})$ is shown in the following way.
 When we apply the BRST charge $Q$ to the product structure, from the Leibniz rule, two terms emerge such that $Q$ acts on the integral domain $\cm$ (as the coboundary operator $\partial$) or on the cohomology numbers $\alpha_i$ ($i=1,2,\ldots,k$). Then, according to the cyclic symmetry and the coboundary formula for the integral domain $\cm$, that is, Eqs. (\ref{eq:cyc1}) and (\ref{eq:cyc2}), the following $A_\infty$ conditions hold.\cite{Nakatsu}
\begin{eqnarray}
\sum_{1\le k,n\le m}(-)^\ast{m}^Q_{m-n}(\alpha_1,\cdots,{m}^Q_{n}(\alpha_k,\cdots,\alpha_{k+n-1}),\cdots,\alpha_{m})=0\;,\label{eq:l}
\end{eqnarray}
where $\ast={\mbox{deg}}(\alpha_1)+\cdots +{\mbox{deg}}(\alpha_{k-1})+k-1$, for all $m\ge1$.

We remark that the morphisms $\alpha=\alpha(\vec{s})$ and the product structures $m_n^Q=m_n^Q(\vec{s})$ ($n=1,2,\ldots$) of $C^Q(\fh)$ also depend on the coordinates $\vec{s}$, just as the objects do.
\section{Equivalence Principle and $A_\infty$ Covariance Principle}
An $A_\infty$ functor between an $A_\infty$ category $C_1$ and another $A_\infty$ category $C_2$, which is different from $C_1$ only in its morphism spaces and product structures---which, in our sense, gives the {\it{coordinate transformation}} between state spaces $V^\nabla|_{S_1}$ and $V^\nabla|_{S_2}$---is defined by the set of an infinite number of maps $\{\cu,\cu_1,\cu_2,\ldots\}$. That is, a map between objects 
\begin{equation}
\cu:C_1\to C_2\;,
\end{equation}
 and the multilinear maps $\cu_k$ with degree $(1-k)$ for all objects $\bc_1,\bc_2,\ldots,\bc_{k+1} \in C_1$ with $k=1,2,\ldots$
\begin{equation}
\cu_k:\bigotimes_{i=1}^k \ch_1(\bc_{i},\bc_{i+1})\to \ch_2(\cu(\bc_1),\cu(\bc_{k+1}))\;,
\end{equation}
 such that for all subsets of the objects $S\subset C_1$ with finite elements
\begin{equation}
\{\cu_k\}_{k\ge1}:\bigoplus_{\bc_1,\bc_2\in S}\ch_1(\bc_1,\bc_2)\to \bigoplus_{\bc_1^\prime,\bc_2^\prime\in \cu(S)}\ch_2(\bc_1^\prime,\bc_2^\prime)\;,
\end{equation}
is an $A_\infty$ morphism between $A_\infty$ algebras of ${\boldsymbol{Z}}$-graded vector spaces with different product structures. 
An $A_\infty$ functor is called an $A_\infty$-{\it{quasi}}-{\it{isomorphism}} when $\cu_1$ gives the quasi-isomorphism between complexes.

In this section, we invoke the minimal model theorem.\cite{Kad,Kajiura} (A {\it{minimal}} $A_\infty$ category is defined such that its product structure ${m}^Q_1$ is trivial: $m_1^Q=0$.)
\begin{Theorem}
For any $A_\infty$ category $C$, there exists a quasi-equivalence $A_\infty$ functor $\cu$ from a minimal $A_\infty$ category $H(C)$ (called the minimal model of $C$) to it. The morphism space of $H(C)$ is the cohomology of the lowest product structure, as defined by Eq.(\ref{eq:low}), of the original $A_\infty$ category $C$.
 \end{Theorem}
 From this statement, the morphism space of the minimal model of $C^Q({\mathfrak{H}})$ consists of {{closed forms}} for the BRST charge $Q$ of the gauged S-duality. 

From the minimal model theorem, the $A_\infty$ functor ${\cal{U}}$ plays a crucial role in revealing the geometrical structure produced by introducing the nonlinear potential to glue together different BRST invariant wave functions $\Psi^Q$ to form the generalized BRST invariant wave function $\Psi^\nabla$.\cite{Konishi} Namely, the {{equivalence principle for all the quantized gauge interactions as Chan-Paton gauge interactions}} holds. In the following, we consider this in more detail. Here, we note that the perturbative structures of the Chan-Paton gauge interactions appear in the {{product structures}} 
\begin{equation}
{m}^Q_n(\alpha_1,\ldots,\alpha_n)\;,\ \ n=1,2,\ldots
\end{equation}
of the $A_\infty$ category, similar to the interactions in the Fukaya category\cite{Fukaya} and quantum cohomology. The statement of this principle is very simple: {\it{the BRST invariant wave function $\Psi^Q$ with nontrivial Chan-Paton interactions (D-particle scattering) cannot be distinguished from the wave function $\Psi^\nabla$ with {{no}} Chan-Paton interaction obtained from $\Psi^Q$ via a certain $A_\infty$ functor.}} That is, {{for given Chan-Paton gauge interactions ${m}^Q_n$ ($n=1,2,\ldots$), describing the D-particle scattering, and for any vacuum $\Psi^\nabla$ for a locality around the coordinates $\vec{s}$, by taking a certain $A_\infty$ functor ${\cal{U}}$, we can completely eliminate these interactions ({\it{as a whole not in each perturbative degree}}), obtaining the free motion
\begin{equation}
{m}^Q_1({\cal{U}}_1(\alpha))=0\;,\label{eq:null2}
\end{equation} under this vacuum in its lowest product structure Eq.(\ref{eq:low}). (Here, we recall that the BRST open string field theory requires the Maurer-Cartan equation for (the equation of motion of) open string field $\alpha$, which ensures the $A_\infty$ associativity of $\alpha$\cite{Kajiura}:
\begin{equation}
\sum_{n=1}^\infty m_n^Q(\alpha,\ldots,\alpha)=0\;,\label{eq:MC}
\end{equation}
and in our claim the summation of all of the interaction terms in it vanishes.
We note that from its definition any $A_\infty$ functor commutes with $m^Q=\sum_{n\ge1}m^Q_n$ of $H(C)$ and $C$.\footnote{In Eq.(\ref{eq:null2}) we use the fact that ${\cal{U}}_1$ commutes with $m_1^Q$ of $H(C)$ and $C$. This fact follows from this definition of an $A_\infty$ functor.\cite{Kajiura}} Thus the quasi-equivalence $A_\infty$ functor in Theorem 1 maps the solutions of Eq.(\ref{eq:MC}) in $H(C)$ to those in $C$.) The existence of such the $A_\infty$ functor $\cu$ is ensured by Theorem 1.

We need to point out two mathematical facts regarding the statement of this equivalence principle. First, the minimal model of an $A_\infty$ category is unique under $A_\infty$ equivalence.\cite{Kad,Kajiura} Second, we assume only one $A_\infty$ product structure $m_n^Q$ ($n=1,2,\ldots$) of $C^Q(\fh)$ defined in Section 2. Due to these two facts, this equivalence principle is mathematically consistent.

 The locality around the coordinates $\vec{s}$ of $\Psi^\nabla$ is an object in $C^Q({\mathfrak{H}})$. Hence, by using a certain coordinate frame of $\vec{s}$, we can locally eliminate the lowest interaction structure ${m}^Q_1$, which results in the renormalizations of higher structures ${m}^Q_n$ ($n=2,3,\ldots$) of the vacuum. (Here, we note again that $m$ indicates not the dynamics but the structure of the interactions.)
 Besides this, the $A_\infty$ covariance principle for the physical substance of the interactions, that is, the quantum mechanical product structures ${m}^Q_n$ ($n=1,2,\ldots$) in $C^Q({\mathfrak{H}})$ for $A_\infty$ functors, is important for the categorification of the theory of gauged S-duality after introducing the nonlinear potential. The well-definedness of this covariance principle is based on the equivalence principle.

The geometry of the ``${\cal{U}}$-manifold'' (referring to Eq.(\ref{eq:trf})), for an arbitrary number $\ell=1,2,\ldots,\infty$,\begin{eqnarray}
\Psi^\nabla|_S&=&\Biggl(\cdots\Biggl(\Biggl(\Psi^Q|_{S_1}\bigcup_{{\cal{U}}_{(1)}} \Psi^Q|_{S_2}\Biggr)\bigcup_{{\cal{U}}_{(2)}}\Psi^Q|_{S_3}\Biggr)\cdots\bigcup_{{\cal{U}}_{(\ell-1)}}\Psi^Q|_{S_{\ell}}\Biggr)\;,\nonumber\\ S&=&\bigcup_{i=1}^{\ell}S_i\;,\label{eq:Uman}
\end{eqnarray}obtained by gluing the restricted vacua via $A_\infty$ functors gives the geometrization of all of the quantized gauge interactions that exist under the vacuum to be considered. In Eq.(\ref{eq:Uman}), the Chan-Paton interactions are created by the glues ${\cal{U}}_{(i)}$ ($i=1,2,\ldots,\ell-1$) due to the existence of the nonlinear second gauge potential in the covariant derivative $\nabla$. We define a (locally $A_\infty$, but in the following we refer to it as just $A_\infty$) category $C^\nabla(\fh)$ with its product structures $m_n^\nabla$ ($n=1,2,\ldots$) such that \begin{eqnarray}{{\bs{m}}^{Q}}&=&(m_n^{Q}(\vec{s}))_n\;,\ \ {{\bs{m}}^{\nabla}}=(m_n^{\nabla}(\vec{s}))_n\;,\nonumber\\ {\bs{m}}^\nabla&=&\Biggl(\cdots\Biggl(\Biggl({\bs{m}}^Q|_{S_1}\bigcup_{{\cal{U}}_{(1)}} {\bs{m}}^Q|_{S_2}\Biggr)\bigcup_{{\cal{U}}_{(2)}}{\bs{m}}^Q|_{S_3}\Biggr)\cdots\bigcup_{{\cal{U}}_{(\ell-1)}}{\bs{m}}^Q|_{S_{\ell}}\Biggr)\;,\end{eqnarray} correspond to the wave function $\Psi^\nabla$ by gluing the patch $A_\infty$ category $C^Q(\fh)$ via the $A_\infty$ functor $\cu$ according to Eq.(\ref{eq:Uman}).

 In the previous paper,\cite{Konishi} by regarding the BRST charge $Q$ as an $\widehat{sl(2,{\bs{R}})}$ invariant Hamiltonian of a system, we showed that for the local wave function $\Psi^Q$, the temporal non-unitary evolution of strings induced by variance of the increment of time (in the sense of a quantum gravity effect), is given by a quasi-equivalence class of projective resolutions ($Q$-complexes) in the derived category with the BRST charge $Q$. By taking the quasi-equivalence class, we remove the ambiguity in the definitions of the states with same variance of the increment of time. Due to our two new principles of equivalence and $A_\infty$ covariance, the refined temporal non-unitary evolution of strings via quantized Chan-Paton interactions is given by an $A_\infty$ quasi-equivalence class of `{\it{projective resolutions}}' on the $A_\infty$ geometry. That is, a quasi-equivalence class of {\it{twisted $A_\infty$ complexes in the derived category $D(C^\nabla(\fh))$ (with a triangulated structure) of the glued $A_\infty$ category $C^\nabla(\fh)$}}. (The difference between the old and the refined model is in their morphism spaces. In the present refined model, by extending the morphism space, we introduce the new degrees of freedom of interaction data via product structures.  Namely, the refined model of the quantum mechanical world is a generalization of the old model in that the former has many body systems (i.e., systems with internal degrees of freedom) as elements.
 When we define the quasi-equivalence class, the refined model has more precision than the old model on this point.) Here, a {\it{twisted $A_\infty$ complex}} is a pair $({\cal{C}},{\cal{Q}})$. ${\cal{C}}$ is an object---a finite formal direct sum of objects of $C^\nabla(\fh)$ with $\bs{Z}$ numbers sliding the grades of objects---of the additive enlargement $S(C^\nabla(\fh))$ of the original glued $A_\infty$ category $C^\nabla(\fh)$, by which we consider the relation systems of multiple D-particle states and transitions between them. ${\cal{Q}}$ is an element of the degree one endomorphism space of the object ${\cal{C}}$, which is the generalization of the BRST charge $Q$, such that the operator ${\cal{Q}}$ is strictly upper triangular in its matrix representation (as explained in the next paragraph) and satisfies the condition for the product structures ${m^{\nabla,S}_n}$ ($n=1,2,\ldots$) of $S(C^\nabla(\fh))$,
\begin{equation}\sum_{n=1}^\infty {m^{\nabla,S}_n}({\cal{Q}},\ldots,{\cal{Q}})=0\;,\label{eq:tw}\end{equation}to which the nilpotency condition, as seen in that on ${m^{\nabla,S}_1}$, is generalized.\cite{Seidel,FOOO}

 We now explain the details of Eq.(\ref{eq:tw}). ${\cal{Q}}$ has a matrix representation labeled by two indices of the slid ${\bs{Z}}$-grading numbers and Eq.(\ref{eq:tw}) is a simultaneous equation for the matrix elements of ${\cal{Q}}$. Then, the graded morphism structure between an object ${\cal{C}}$ of $S(C^\nabla(\fh))$ and itself, as the direct formal sum of the objects in $C^\nabla(\fh)$ with grading, appearing in the product of matrices in ${\cal{Q}}$ produces the higher terms ${m^{\nabla,S}_k}$ ($k\ge2$). But, these higher terms have no meaning that is essentially different from the ${m^{\nabla,S}_1}$ term in the $A_\infty$ category $C^\nabla(\fh)$. Actually, on $C^\nabla(\fh)$, the condition in Eq.(\ref{eq:tw}) is just ${m^\nabla_1}({\cal{Q}})=0$. In this sense, ${\cal{Q}}$ is the generalization of $Q$ to the relation system of D-particle states. Upper triangularity of ${\cal{Q}}$ means that Eq.(\ref{eq:tw}) is a finite sum.

 The category of twisted $A_\infty$ complexes with the morphism spaces in $S(C^\nabla(\fh))$ has an $A_\infty$ structure, and we can define the derived category of the original $A_\infty$ category by the degree zero cohomology of the $A_\infty$ category of twisted $A_\infty$ complexes.\cite{Seidel,FOOO} As noted from Eq.(\ref{eq:tw}), the operator ${\cal{Q}}$ reflects the interaction structures of D-particle states, in the given relation system of D-particle states ${\cal{C}}$ with its graded morphism structure, that is, the product structures ${m^{\nabla,S}_n}$ ($n=1,2,\ldots$) in $S(C^\nabla(\fh))$ generated by the nonlinear second gauge potential. Regarding this point, we note that our $A_\infty$ category is {\it{background independent}} and all of space-time and all matter consist of its objects. There is no isolated and completed object, and any object has morphisms between other objects and belongs to the (nontrivial) relation systems, which are explained above.

\section{Summary}
Continuing from the author's previous work\cite{Konishi}, we investigated the principles of equivalence and $A_\infty$ covariance of the theory of gauged S-duality via the D-particle field category by using the framework of the geometry of geodesics and their intersections on the upper half plane as a toy model. The equivalence principle is based on the minimal model theorem of the $A_\infty$ category.\cite{Kajiura} The infinitesimal descriptions of $\Psi^\nabla$ under the nonlinear second gauge potential are classified as the different classes of degenerations of geodesics drawn on an upper half plane. Using this fact, we produced a toy model of the $A_\infty$ category of the theory of gauged S-duality. On the basis of these two principles, we describe all the quantized Chan-Paton gauge interactions in terms of the geometry of the ${\cal{U}}$-manifold $\Psi^\nabla$, or equivalently $C^\nabla(\fh)$ defined according to Eq.(\ref{eq:Uman}). Within this manifold $\Psi^\nabla$, or equivalently $C^\nabla(\fh)$, the corresponding notion to a Euclidean patch is the local wave function $\Psi^Q$, or equivalently the $A_\infty$ category $C^Q(\fh)$ under the local description, and gluing is done by an $A_\infty$ functor ${\cal{U}}$. The temporal non-unitary evolution of strings is given by quasi-equivalence class of twisted $A_\infty$ complexes, that is, the generalization of the quasi-equivalence class of complexes of D-particle states to that of the relation systems of D-particle states.


\begin{thebibliography}{99}
\bibitem{Konishi}E. Konishi, Int. J. Mod. Phys. {\bf{A 26}}, 4785 (2011), arXiv:1001.3382 [hep-th].
\bibitem{Sen}A. Sen, Int. J. Mod. Phys. {\bf A 9}, 3707 (1994), arXiv:hep-th/9402002.
\bibitem{Schwarz}J. H. Schwarz, Phys. Lett. {\bf{B 360}}, 13 (1995), arXiv:hep-th/9508143.
\bibitem{Witten}E. Witten, 
{Nucl. Phys.} {\bf{B}} {\bf{443}}, 85 (1995), arXiv:hep-th/9503124.
\bibitem{Polchinski}J. Polchinski, 
{Phys. Rev. Lett.} {\bf{75}}, 4724 (1995), arXiv:hep-th/9510017.
\bibitem{Y1}T. Yoneya, JHEP {\bf{12}}, 28 (2005), arXiv:hep-th/0510114.
\bibitem{Y2}T. Yoneya, Prog. Theor. Phys. {\bf{118}}, 135 (2007), arXiv:0705.1960 [hep-th].
\bibitem{Y3}T. Yoneya, Int. J. Mod. Phys. A {\bf{23}}, 2343 (2008), arXiv:0804.0297 [hep-th].
\bibitem{K}E. Konishi, Prog. Theor. Phys. {\bf{121}}, 1125 (2009), arXiv:0902.2565 [hep-th].
\bibitem{BFSS}T. Banks, W. Fischler, S. H. Shenker and L. Susskind, Phys. Rev. D {\bf{55}}, 5112 (1997), arXiv:hep-th/9610043.
\bibitem{IKKT}N. Ishibashi, H. Kawai, Y. Kitazawa and A. Tsuchiya, Nucl. Phys. B {\bf{498}}, 467 (1997), arXiv:hep-th/9612115.
 \bibitem{Kad}T. V. Kadeishvili, Soobshch. Akad. Nauk. Gruzin. SSR. {\bf{108}}, 249 (1982).
 \bibitem{Kajiura}H. Kajiura, Rev. Math. Phys. {\bf{19}}, 1 (2007), arXiv:math.QA/0306332.
\bibitem{Fukaya}K. Fukaya, Morse homotopy, $A_\infty$ category and Floer homologies, in {\it{Proc. GARC Workshop on Geometry and Topology}}, Seoul, 1993, Lecture Notes Series, Vol. 18, ed. H. J. Kim (Seoul National University, 1993).
\bibitem{planar}E. Konishi, Int. J. Mod. Phys. {\bf{A 22}}, 5351 (2007), arXiv:0707.0387 [hep-th].
\bibitem{PA}K. Bardakci and H. Ruegg, Phys. Rev. {\bf{181}}, 1884 (1969).
\bibitem{Nakatsu}T. Nakatsu, Nucl. Phys. B {\bf{642}}, 13 (2002), arXiv:hep-th/0105272.
 \bibitem{Seidel}P. Siedel, Homological mirror symmetry for the quartic surface, preprint, arXiv:math/0310414.
 \bibitem{FOOO}K. Fukaya, Y.-G. Oh, H. Ohta and K. Ono, {\it{Lagrangian Intersection Floer theory ---Anomaly and obstructions---}}, Part I and Part II, AMS/IP Studies in Advanced Mathematics, vol {\bf{46.1}} and vol {\bf{46.2}}, Amer. Math. Soc./International Press, 2009.
\end{thebibliography}
\end{document}